\documentclass[useAMS,usenatbib,usegraphicx]{mn2e}
\usepackage{float}
\newcommand{\bugref}{\bibitem[\protect\citeauthoryear{dummy }{1893}]{dum}}

\title[Transverse Faraday-Rotation Gradients Across AGN Jets]
{Transverse Faraday-Rotation Gradients Across the Jets of 15 Active 
Galactic Nuclei}

\author[D. C. Gabuzda, S. Knuettel \& B. Reardon]
{D. C. Gabuzda, S. Knuettel \& B. Reardon \\
Physics Department, University College Cork, Cork, Ireland} 

\begin{document}

\date{}
\pagerange{\pageref{firstpage}--\pageref{lastpage}} \pubyear{2013}
\maketitle
\label{firstpage}
\begin{abstract}
The presence of a helical magnetic field threading the jet of an
Active Galactic Nucleus (AGN) should give
rise to a gradient in the observed Faraday rotation measure (RM) across 
the jet, due to the associated systematic change in the line-of-sight 
magnetic field.  Reports of observations of transverse RM 
gradients across AGN jets have appeared in the literature starting 
from 2002, but concerns were raised about the resolution required for 
these gradients to be reliable, and there was a lack of a full 
understanding of the best approach to accurate estimation of the 
uncertainties of local RM values.  These questions have now been 
resolved by recent Monte Carlo simulations carried out by various
groups, enabling both a verification of previously published results and 
reliable analyses of new data. We consider here RM gradients across 
the jet structures of 15 AGN, some previously published in the refereed 
literature but without a correct and complete error analysis, and 
some published for the first time here, all of which have monotonic 
transverse RM gradients  with significances of at least $3\sigma$. 
\end{abstract}

\section{Introduction}

The radio emission associated with Active Galactic Nuclei (AGNs)
is synchrotron emission, which can be linearly polarized up to about 
75\% in optically thin regions, where the polarization angle
$\chi$ is orthogonal to the projection of the magnetic field 
{\bf B} onto the plane of the sky, and 
up to 10--15\% in optically thick regions, where $\chi$ is parallel to
the projected {\bf B} (Pacholczyk 1970). Linear polarization measurements thus
provide direct information about both the degree of order and the
direction of the {\bf B} field giving rise to the observed synchrotron
radiation.  

Multi-frequency Very Long Baseline Interferometry (VLBI) polarization
observations also provide  information about the parsec-scale 
distribution of the spectral index (optical depth) of the emitting 
regions, as well as Faraday rotation occurring between the source 
and observer.  Faraday rotation of the plane of linear  polarization 
occurs during the passage of the associated electromagnetic wave 
through a region with free electrons and a {\bf B} field with a 
non-zero component along the line of sight. When the Faraday rotation 
occurs outside the emitting region in regions of non-relativistic 
(``thermal'') plasma, the amount of rotation is given by
\begin{eqnarray}
           \chi_{obs} - \chi_o = 
\frac{e^3\lambda^{2}}{8\pi^2\epsilon_om^2c^3}\int n_{e} 
{\mathbf B}\cdot d{\mathbf l} \equiv \textrm{RM}\lambda^{2}
\end{eqnarray}
where $\chi_{obs}$ and $\chi_o$ are the observed and intrinsic 
polarization angles, respectively, $-e$ and $m$ are the charge and 
mass of the particles giving rise to the Faraday rotation, usually 
taken to be electrons, $c$ is the speed of light, $n_{e}$ is the 
density of the Faraday-rotating electrons, $\mathbf{B}$ is the
magnetic field, $d\mathbf{l}$ is an element along the line of 
sight, $\lambda$ is the observing wavelength, and RM (the coefficient of 
$\lambda^2$) is the Rotation Measure (e.g., Burn 1966).  Simultaneous 
multifrequency observations thus allow the determination of both 
the RM, which carries information about the electron density and 
the line-of-sight {\bf B} field in the region of Faraday rotation, 
and $\chi_o$, which carries information about the intrinsic {\bf B}-field 
geometry associated with the source projected onto the plane of the sky.

As was pointed out by Blandford (1993), the presence of a helical 
{\bf B} field threading the jet of an AGN should give rise to a 
gradient in the observed RM across the jet, due to the associated 
systematic change 
in the line-of-sight {\bf B} field.  Further, such fields 
would come about in a natural way as a result of the ``winding up'' of 
an initial ``seed'' field by the rotation of the central accreting 
objects (e.g. Nakamura et al. 2001; Lovelace et al. 2002).

The first report of an actual detection of such a transverse RM gradient 
was made by Asada et al. (2002), for the VLBI jet of 3C273; this result 
was later confirmed by Zavala \& Taylor (2005) and Hovatta et al. 
(2012).  Transverse RM gradients  were subsequently reported across 
the parsec-scale jets of a number of other AGN (e.g., 
Gabuzda et al. 2004, 2008; Asada et al. 2008, 2010; Kharb et al. 2009; 
Mahmud et al. 2009; Croke et al. 2010), and
interpreted as reflecting the systematic change in the line-of-sight 
component of a toroidal or helical jet {\bf B} field across the jets.
 
\begin{center}
\begin{table*}
\begin{tabular}{c|c|c|c|c|c|c|c}
\hline
\multicolumn{8}{c}{Table 1: Source properties}\\
Source & Redshift & Optical & pc/mas &  Ref & Integrated RM & Ref & Original RM   \\ 
       &          &  ID &      &      & (rad/m$^2$)  &     &  Map Ref \\\hline
0256+075  & 0.89 & BL & 7.78   & S89, KS90  & $-13$  &  P  &  *     \\
0355+508  & 1.52 &  Q & 8.54   &   M  & $-17$  &  R  &  ZT    \\
0735+178  & 0.45 & BL & 5.73   &   M  & $+9$   &  R  &  G2    \\
0745+241  & 0.41 & HPQ& 5.42   &   M  & $+21$  &  P  &  G1    \\
0748+126  & 0.89 & Q & 7.78   &   M  & $+14$  &  T  &  ZT    \\
0820+225  & 0.95 & BL   & 7.93   & S93, KS90  & $+81$  &  P  &  G1    \\
0823+033  & 0.50 & BL & 6.12   &   M  & $+1$   &  P  &  *     \\
1156+295  & 0.72 & HPQ & 7.25   &   M  & $-32$  &  T  &  G2    \\
1219+285  & 0.10 & BL & 1.87   &   M  &  $-1$  &  R  &  *  \\
1334$-$127& 0.54 & HPQ & 6.33   &   M  & $-23$  &  P  &  *     \\
1652+398  & 0.034& BL & 0.66   &   M  & $+42$  &  R  &  G1    \\
1749+096  & 0.32 & BL & 4.64   &   M  & $+94$  &  P  &  G2\\
1807+698  & 0.051& BL & 0.98   &   M  & $+11$  &  T  &  G1    \\
2007+777  & 0.34 & BL & 4.83   &   M  & $-20$  &  R  &  *     \\ 
2155$-$152& 0.67 & HPQ & 7.02   &   M  & $+19$  &  P  &  *   \\ \hline
\multicolumn{8}{l}{BL = BL Lac object; Q = Quasar with no optical 
polarization data; HPQ = Quasar with}\\
\multicolumn{8}{l}{fractional linear polarization in the optical above 
3\% on at least one occasion; M = MOJAVE }\\
\multicolumn{8}{l}{website; S89 = Stickel et al. 1989; 
S93 = Stickel et al. 1993; KS90 = K\"{u}hr \& Schmidt 1990;}\\
\multicolumn{8}{l}{R = Rusk 1988; P = Pushkarev 2001; T = Taylor et al. 2009;
ZT = Zavala \& Taylor (2004);}\\
\multicolumn{8}{l}{G1 = Gabuzda et al. (2004); G2 = Gabuzda et al. (2008); * = 
Not previously published in the}\\
\multicolumn{8}{l}{refereed literature}\\
\end{tabular}
\end{table*}
\end{center}


\begin{center}
\begin{table*}
\begin{tabular}{c|c|c|c|c|c|c|c}
\hline
\multicolumn{8}{c}{Table 2: Map properties}\\
Source & Figure & Freq &  Peak  & Lowest contour & BMaj & BMin & BPA \\ 
      &    & (GHz)   &  (Jy) &  (\%) & (mas) & (mas) & (deg) \\\hline
0256+075   & 1a & 4.6 &  0.38  & 0.25 & 3.35  & 1.69  & -3.2    \\
0355+508   & 3a & 8.1 &  4.42  & 0.50 & 1.98  & 0.94  &  21.7   \\
0748+126   & 3b & 8.1 &   0.90 & 0.25 &  2.25 &  0.86 &  -7.7   \\
0823+033   & 1b & 5.0 &  1.16  & 0.25 &  3.67 &  1.64 &  -0.1   \\
1219+285   & 1c & 5.0 &  0.28  & 0.50 &  3.68 & 2.58  & -13   \\
1334$-$127 & 2a & 4.6 &  3.30  & 0.25 &  3.76 & 1.49  &  -0.9   \\
1334$-$127 & 2d & 4.6 &  3.27  & 0.25 &  2.00 & 2.00  &  0     \\
2007+777   & 1d & 5.0 &  0.95  & 0.25 &  1.82 & 1.66  & -20.4   \\ 
2155$-$152 & 2b,c & 4.6 & 1.02   & 0.25 &  4.24 & 1.68  & -1.7  \\ 
2155$-$152 & 2e & 4.6 & 1.01   & 0.50 &  2.66 & 2.66  &  0    \\ \hline
\end{tabular}
\end{table*}
\end{center}

However, there were three main difficulties with these measurements, which
led to skepticism about these results among some researchers.  One was
that it seemed non-intuitive that it could be possible to detect these
RM gradients when the intrinsic widths of the jets were sometimes
appreciably smaller than the widths of the beams with which they were
observed; this was expressed by Taylor \& Zavala (2010) through their
proposed criterion that an observed transverse RM gradient must have a
width of at least three ``resolution elements'' (usually taken to mean
three beamwidths) in order to be considered reliable. This criterion 
was presented without justification, but if it was correct, it would
invalidate nearly all the previously published reports of transverse
RM gradients, as well as make it very difficult to identify new cases.

The second difficulty was that determining the significance of observed
gradients required reasonably accurate estimation of the uncertainties
in the RM value being compared. The standard practice at the time was
to assign an uncertainty to the Stokes $Q$ and $U$ values in individual
pixels equal to the off-source rms in the corresponding $Q$ and $U$
images, $\sigma_{Q,rms}$ and $\sigma_{U,rms}$, then propagate these 
uncertainties to determine the corresponding
uncertainties in the polarization angles $\chi = 0.5 \arctan U/Q$; however,
this approach had never been tested. In addition, attempts were made to
assign uncertainties to RM values averaged over several pixels based on
the standard deviation of the values being averaged, but this ignored the
presence of correlations in the values measured in neaby pixels due to
convolution with the CLEAN beam used. Further, residual instrumental
polarizations remaining in the polarization data after calibration could
potentially add additional uncertainty to the measured polarization 
angles, and this had not been taken into account.

A third difficulty was that there can, in some cases, be small but
significant relative shifts between the polarization angle images 
obtained at different frequencies, due to the frequency dependence of
the position of the VLBI core (Blandford \& K\"{o}nigl 1979). This led
to concerns about whether apparent Faraday rotation gradients could arise
due to incorrect relative alignment of the polarization images. The necessary
shifts can be determined and corrected for, but this had not been done
in a number of studies.

Thus, there was a need to clarify whether there was a minimum resolution
required to reliably detect Faraday-rotation structure, to identify
a suitable approach for accurately estimating uncertainties in quantities 
at individual locations in VLBI images and to ensure that any small relative
shifts between the polarization angle images used to construct the RM
maps were correctly taken into account (or that they were small enough for
their effect on the RM images to be negligible).

A first important step was taken by Hovatta et al. (2012), who carried
out Monte Carlo simulations based on realistic ``snapshot'' baseline coverage
for VLBA observations at 7.9, 8.4, 12.9 and 15.4~GHz, aimed at investigating
the statistical occurrence of spurious RM gradients across jets with
intrinsically constant polarization profiles. Inspection of the
right-hand panel of Fig. 30 of Hovatta et al. (2012) shows that the
fraction of spurious $3\sigma$ gradients was no more than about 1\%,
even for the smallest observed RM-gradient widths they considered,
about 1.4 beamwidths. Relatively few $2\sigma$ gradients were also
found, although this number reached about 7\% for observed jet widths of about
1.5 beamwidths; nevertheless, Hovatta et al. (2012) point out that
$2\sigma$ gradients are potentially also of interest if confirmed over
two or more epochs.

The results of Hovatta et al. (2012) have also now been confirmed by
similar Monte Carlo simulations carried out by Algaba (2013) for simulated
data at 12, 15 and 22~GHz and by Murphy \& Gabuzda (2013) for simulated
data at 1.38, 1.43, 1.49 and 1.67~GHz and at the same frequencies as
those considered by Hovatta et al. (2012).  The 1.38-–1.67~GHz frequency range
considered by Murphy \& Gabuzda (2013) yielded a negligible number of
spurious $3\sigma$ gradients (and fewer than 1\% spurious $2\sigma$
gradients), even for observed jet widths of only 1 beamwidth (i.e., for
poorly resolved jets).

Two more sets of Monte Carlo simulations based on realistic snapshot VLBA
baseline coverage adopted a complementary approach: instead
of considering the occurrence of spurious RM gradients across jets with
constant polarization, they considered simulated jets with various widths
and with transverse RM gradients of various strengths, convolved with
various size beams. Mahmud et al. (2013) carried out such simulations for
4.6, 5.0, 7.9, 8.4, 12.9 and 15.4~GHz VLBA data, and Murphy \&
Gabuzda (2013) for 1.38, 1.43, 1.49 and 1.67~GHz VLBA data.
These simulations clearly showed that, with realistic noise and baseline
coverage, the simulated RM gradients could remain clearly visible, even
when the jet width was as small as 1/20 of a beam width. 

All these simulations clearly demonstrate that the width spanned
by an RM gradient is not a crucial criterion for its reliability,
at least down to 1/20 of a beam width.  This
counterintuitive result essentially comes about because polarization is 
a vector quantity, while the intensity is a scalar. A difference in the
polarization angles across the jet can be detected in situations where
intensity structure could not be. Alternatively, thinking of the polarization
as being composed of Stokes $Q$ and $U$, this enhanced sensitivity to
closely spaced structures comes about because both $Q$ and $U$ can be
positive or negative. It is important to bear in mind that we are not
speaking here of being able to accurately deconvolve the observed
RM profiles to determine the intrinsic transverse RM structure --- only of
the ability to detect the presence of a systematic transverse RM
gradient.

Another key outcome of the Monte Carlo simulations of Hovatta et al. (2012)
is an empirical formula that can be used to estimate the uncertainties
in intensity (Stokes $I$, $Q$ or $U$) images, including the uncertainty due to
residual instrumental polarization (``D-terms'') that has been incompletely
removed from the visibility data. In regions of source emission where the
contribution of the residual instrumental polarization is negligible,
the typical uncertainties in individual pixels are approximately 1.8
times the rms deviations of the flux about its mean value far from regions
of source emission, $\sigma_{rms}$. This is roughly a factor of two greater 
than the uncertainties that have been assigned to intensities measured
in individual pixels as standard practice in past studies, indicating 
that the past uncertainties have been somewhat underestimated. These 
results have recently been
confirmed by the Monte Carlo simulations of Coughlan (2014), which
likewise indicate that the typical intensity uncertainties in individual
pixels are of order twice the off-source rms, with significant pixel-to-pixel
variations in the uncertainties appearing in the case of well resolved
sources.

New Faraday-rotation analyses based on the error formulation of
Hovatta et al. (2012), focusing on monotonicity and a significance of
at least $3\sigma$ as the key criteria for reliability of observed
transverse RM gradients, and ensuring that the RM images analyzed are not
significantly affected by relative shifts between the polarization-angle 
images at different frequencies have begun to appear 
(Mahmud et al. 2013, Gabuzda et al. 2014a, 2014b). 

There is also a need to verify the reliability of previously published 
results. In the current
paper, we present the results of new analyses of the 7 RM maps previously 
published by Gabuzda et al. (2004, 2008), and confirm 
that the previously reported RM gradients are significant. We also report 8 
new cases of monotonic, statistically significant transverse RM gradients 
across AGN jets, based on both published maps and maps not previously
published in the refereed literature, constructed using a variety of datasets 
with three to seven frequencies. 

\begin{figure*}
\begin{center}
\includegraphics[width=.28\textwidth,angle=90]{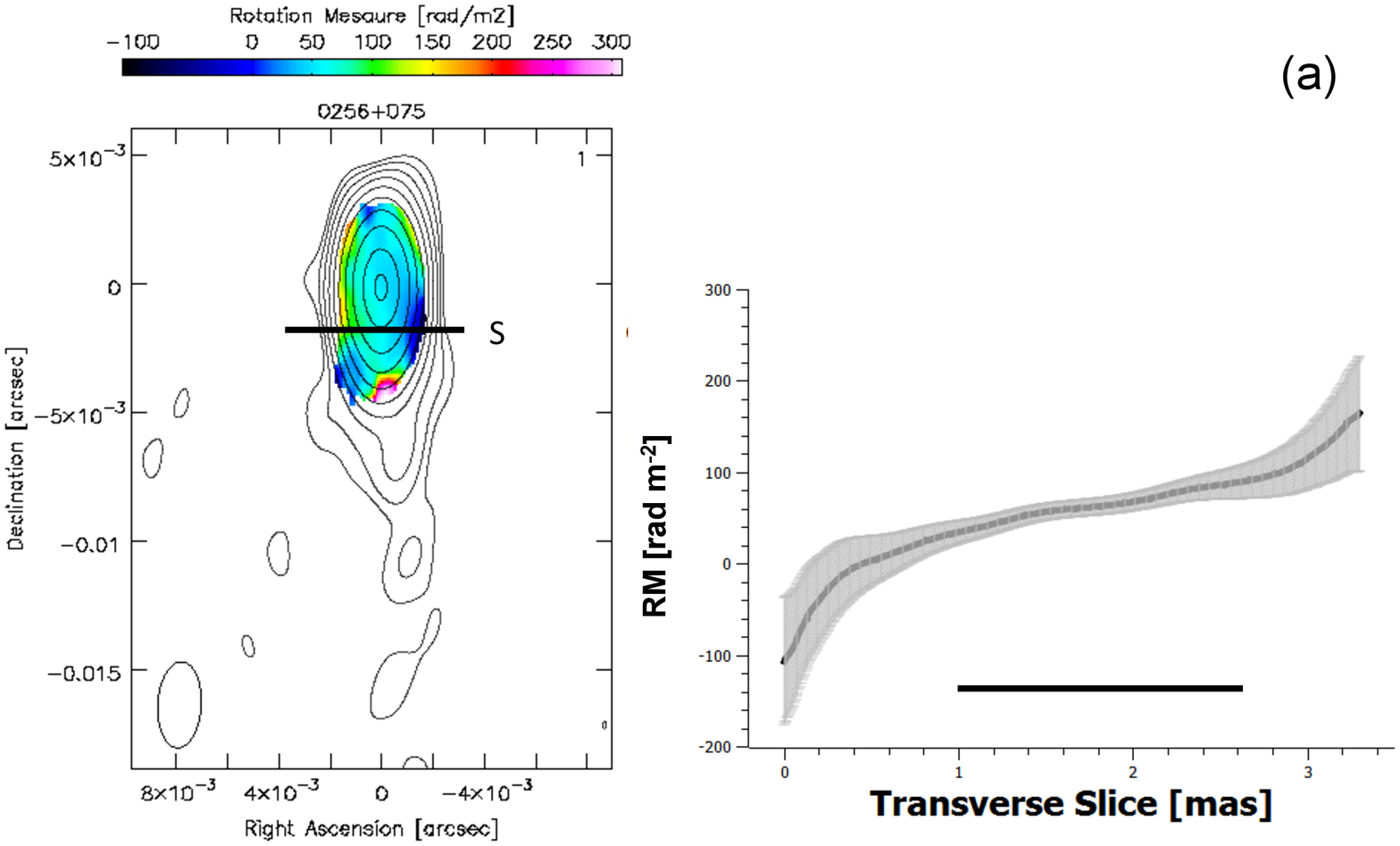}
\includegraphics[width=.28\textwidth,angle=90]{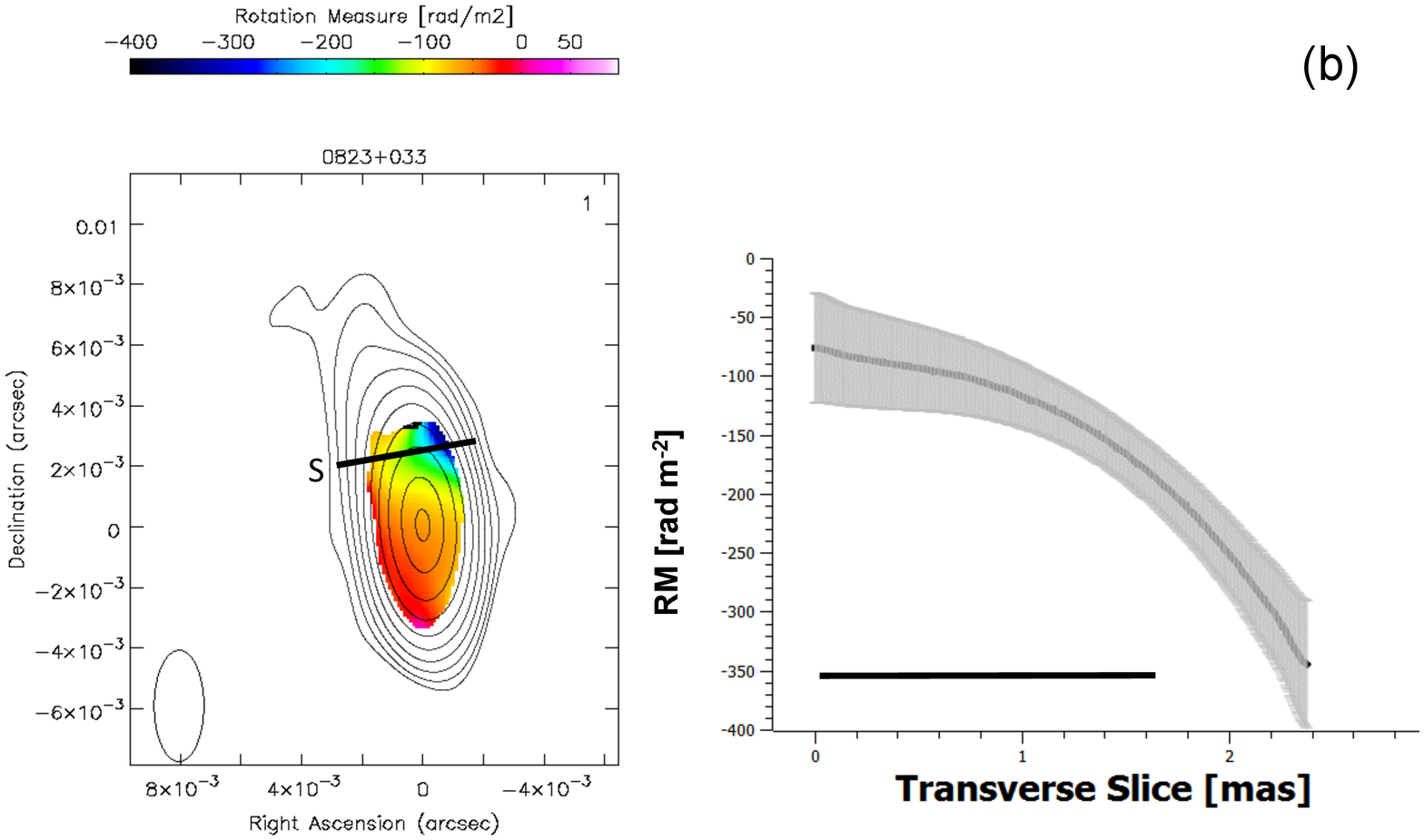}

\vspace*{0.8cm}
\includegraphics[width=.28\textwidth,angle=90]{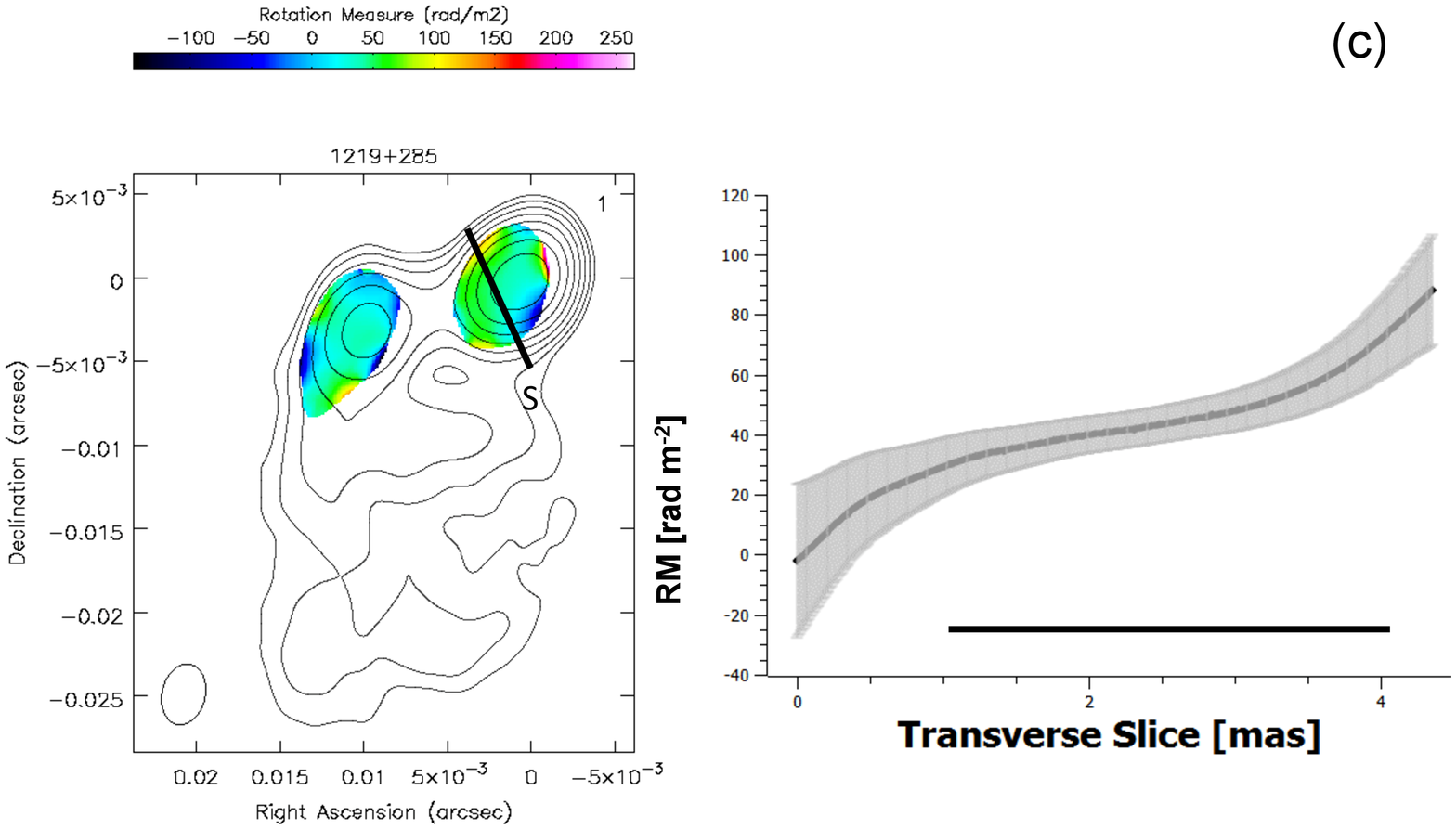}
\includegraphics[width=.28\textwidth,angle=90]{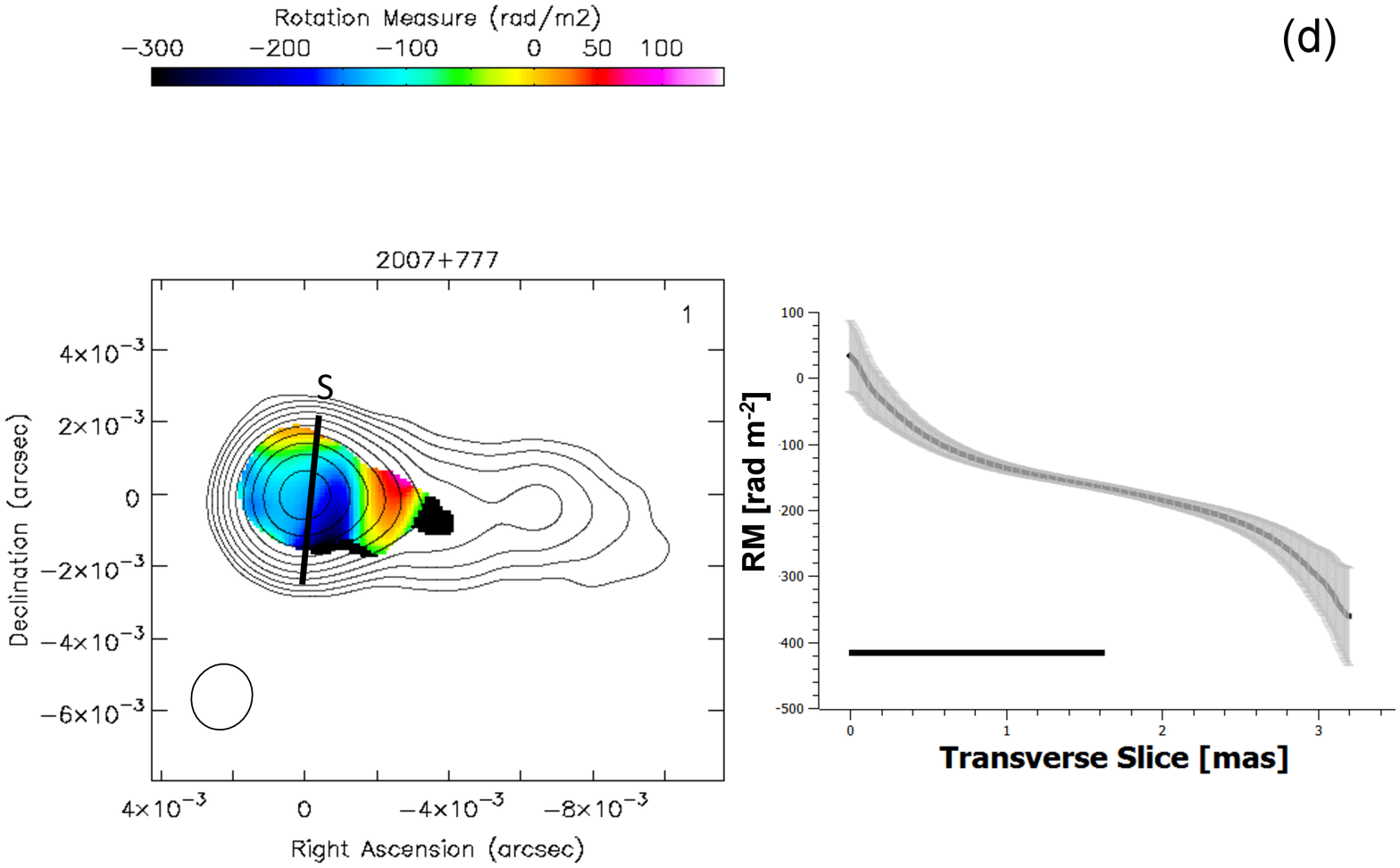}
\end{center}
\caption{(a) 4.6-GHz intensity map of 0256+075 and 5.0-GHz intensity
maps of (b) 0823+033, (c) 
1219+285, and (d) 2007+777 with the corresponding RM distributions 
superposed (left panels).  The lines drawn 
across the RM distributions show the locations of the RM slices shown in the 
corresponding right-hand panels; the letter ``S'' at one end of these 
lines marks the side corresponding to the starting point for the slice. 
The bold horizontal bars shown with the slice profiles indicate the beam 
full widths at half maximum in the direction of the slices, which are 
(a) 1.7~mas, (b) 1.65~mas, (c) 2.7~mas and (d) 1.7~mas. }
\label{fig:rmmaps5-1}
\end{figure*}

\begin{figure*}
\begin{center}
\vspace*{0.2cm}
\includegraphics[width=.30\textwidth,angle=0]{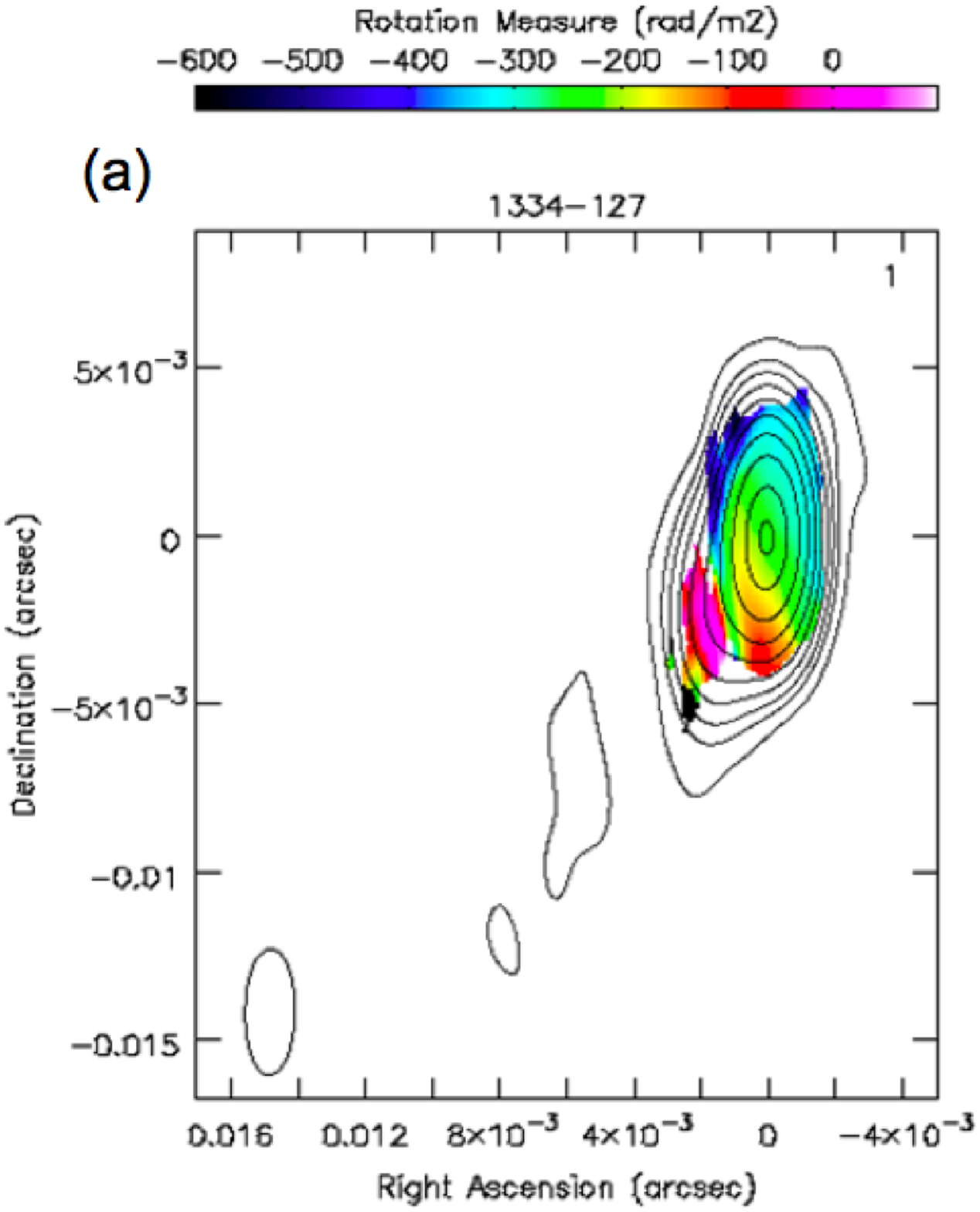}
\includegraphics[width=.28\textwidth,angle=0]{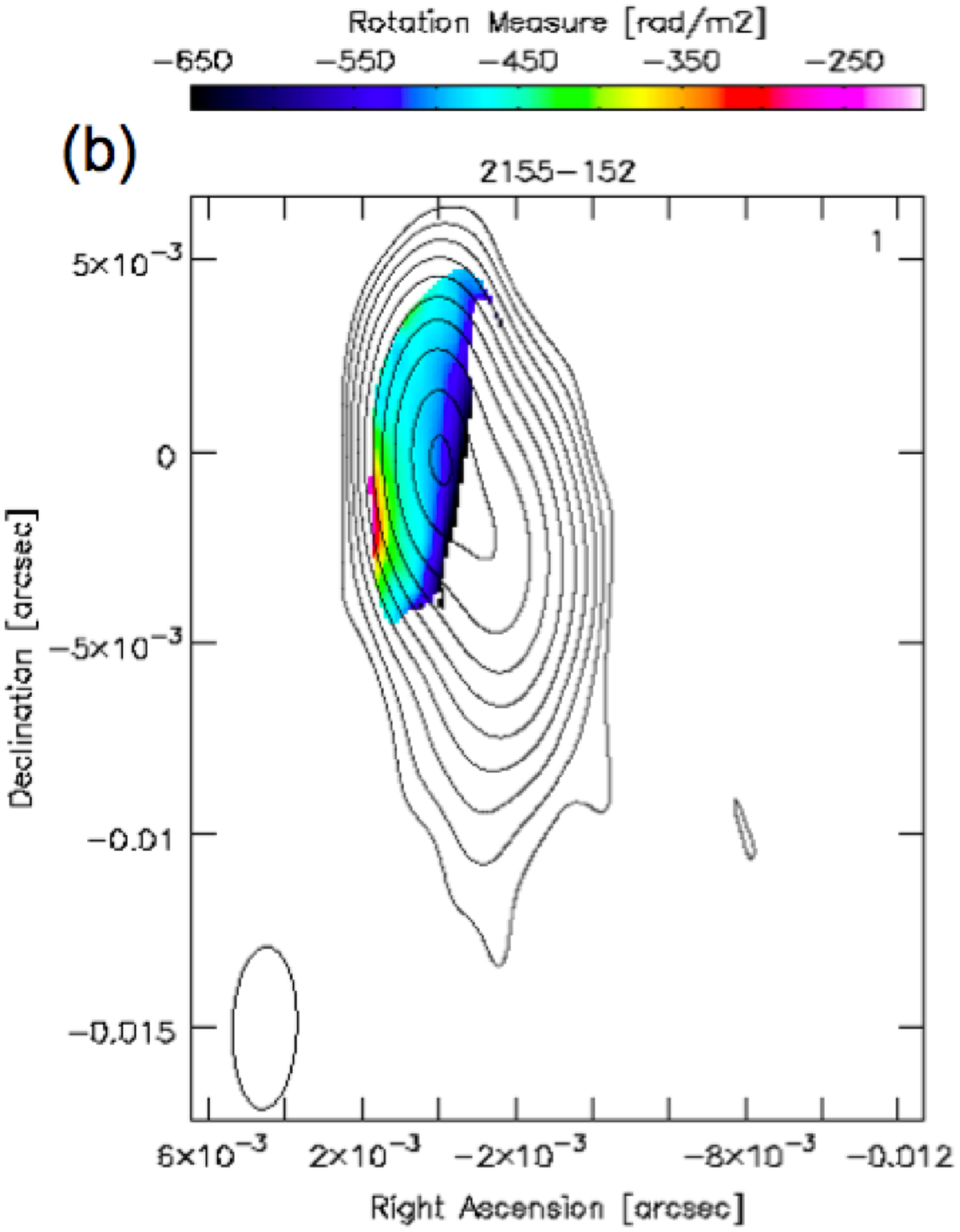}
\includegraphics[width=.28\textwidth,angle=0]{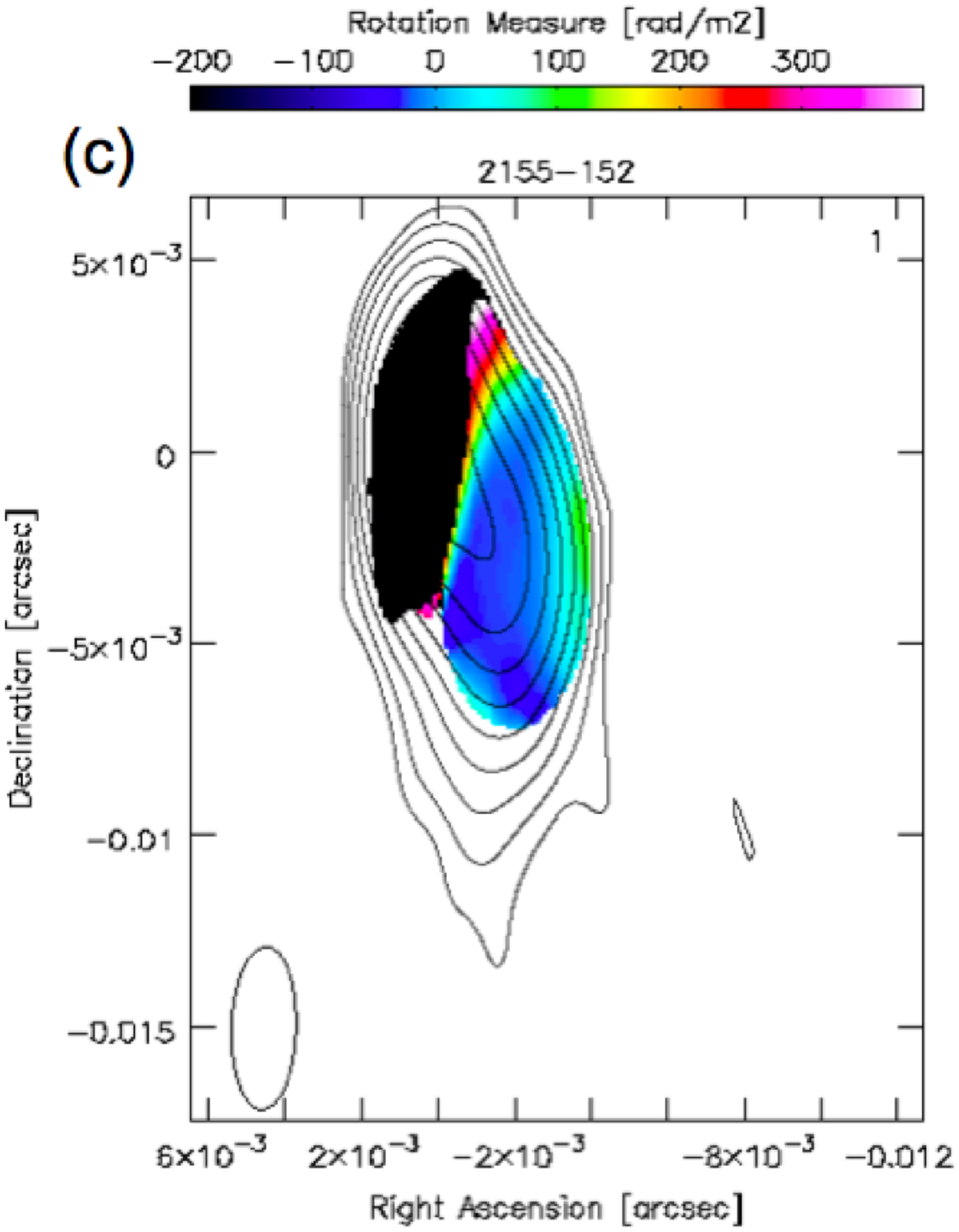}

\vspace*{0.5cm}
\includegraphics[width=.35\textwidth,angle=90]{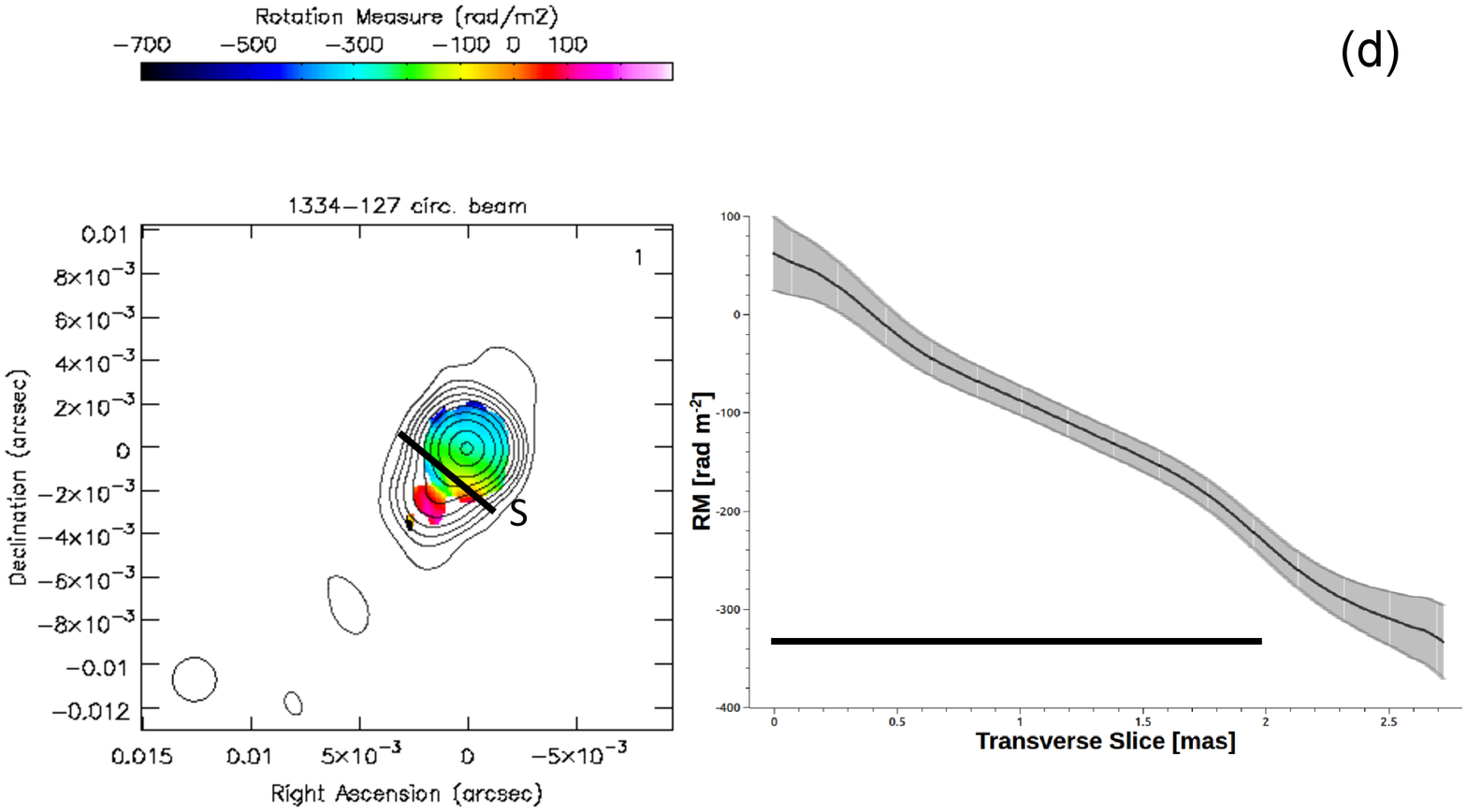}

\vspace*{0.5cm}
\includegraphics[width=.35\textwidth,angle=90]{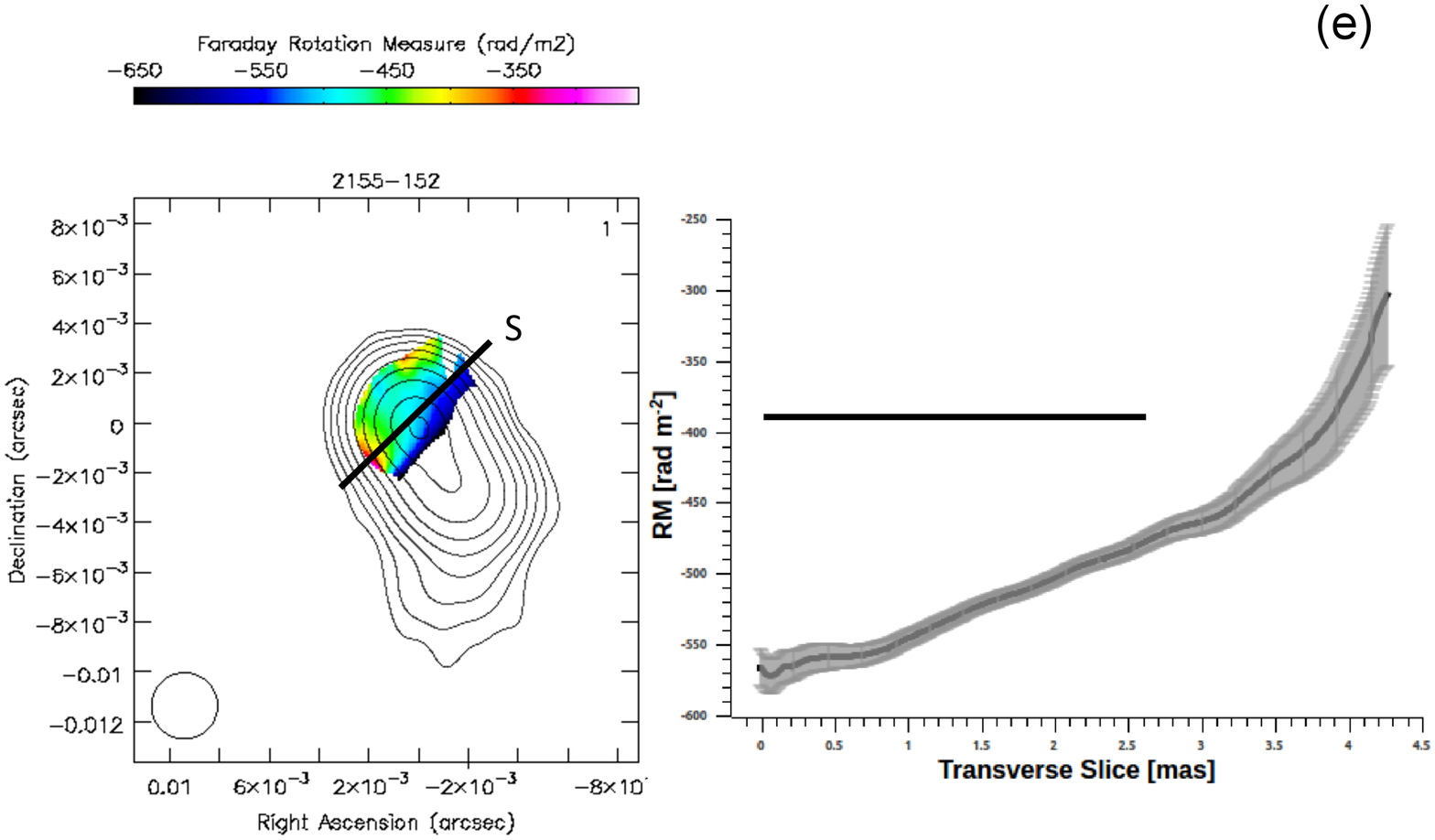}
\end{center}
\caption{4.6-GHz intensity maps of
(a) 1334$-$127 (intrinsic beam), (b) 2155$-$152 core region (intrinsic beam), 
(c) 2155$-$152 jet (intrinsic beam), (d) 1334$-$127 (circular beam) 
(e) 2155$-$152 core region (circular beam). The lines drawn across the 
superposed RM distributions in
panels (d) and (e) show the locations of the RM slices shown in the
corresponding right-hand panels; the letter ``S'' at one end of these
lines marks the side corresponding to the starting point for the slice.
The bold horizontal bars shown with the slice profiles indicate the beam 
full widths at half maximum in the direction of the slices, which are 
(d) 2.0~mas and (e) 2.7~mas. }
\label{fig:rmmaps5-2}
\end{figure*}

\begin{figure*}
\begin{center}
\includegraphics[width=.30\textwidth,angle=90]{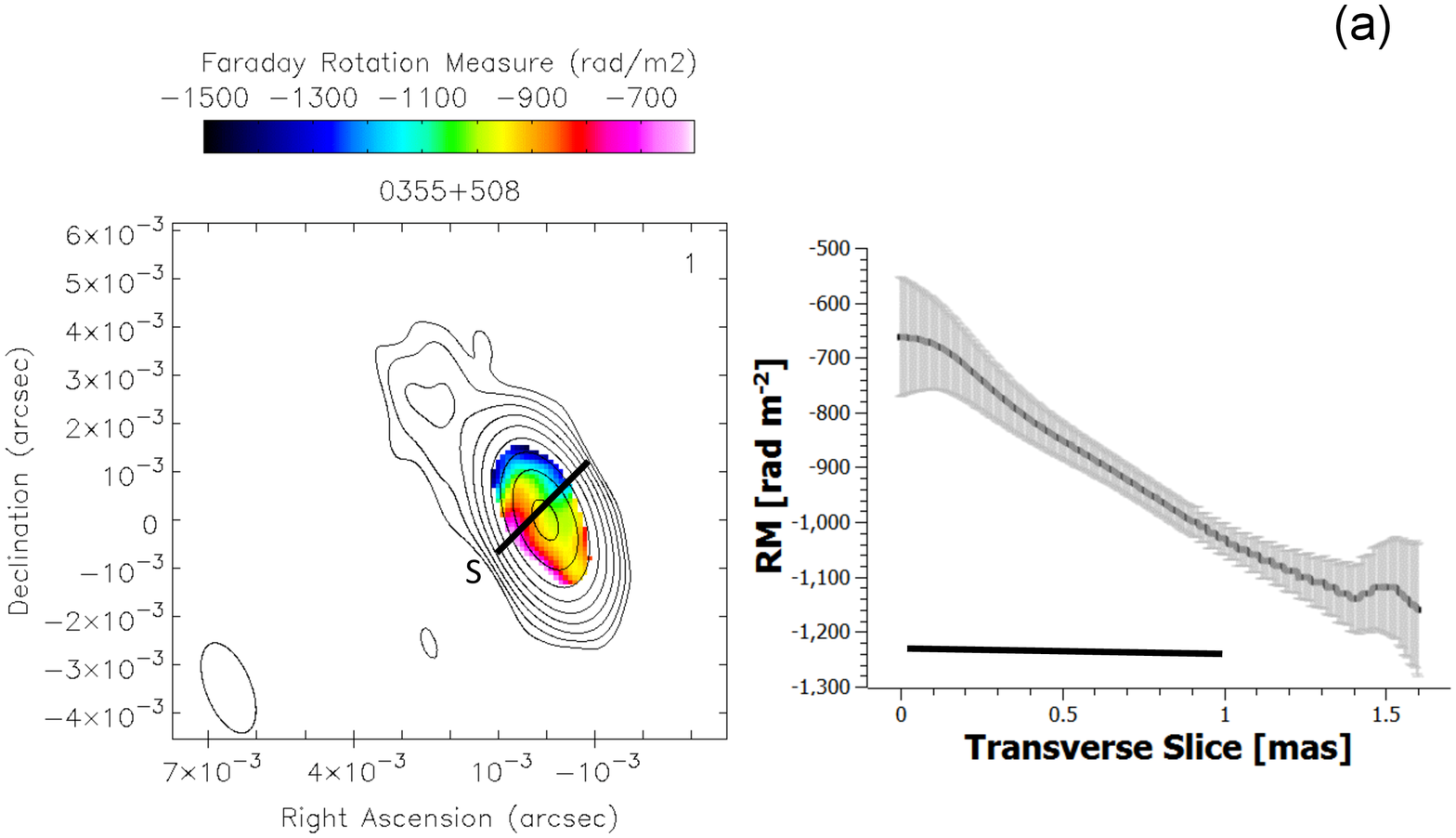}

\vspace*{0.5cm}
\includegraphics[width=.30\textwidth,angle=90]{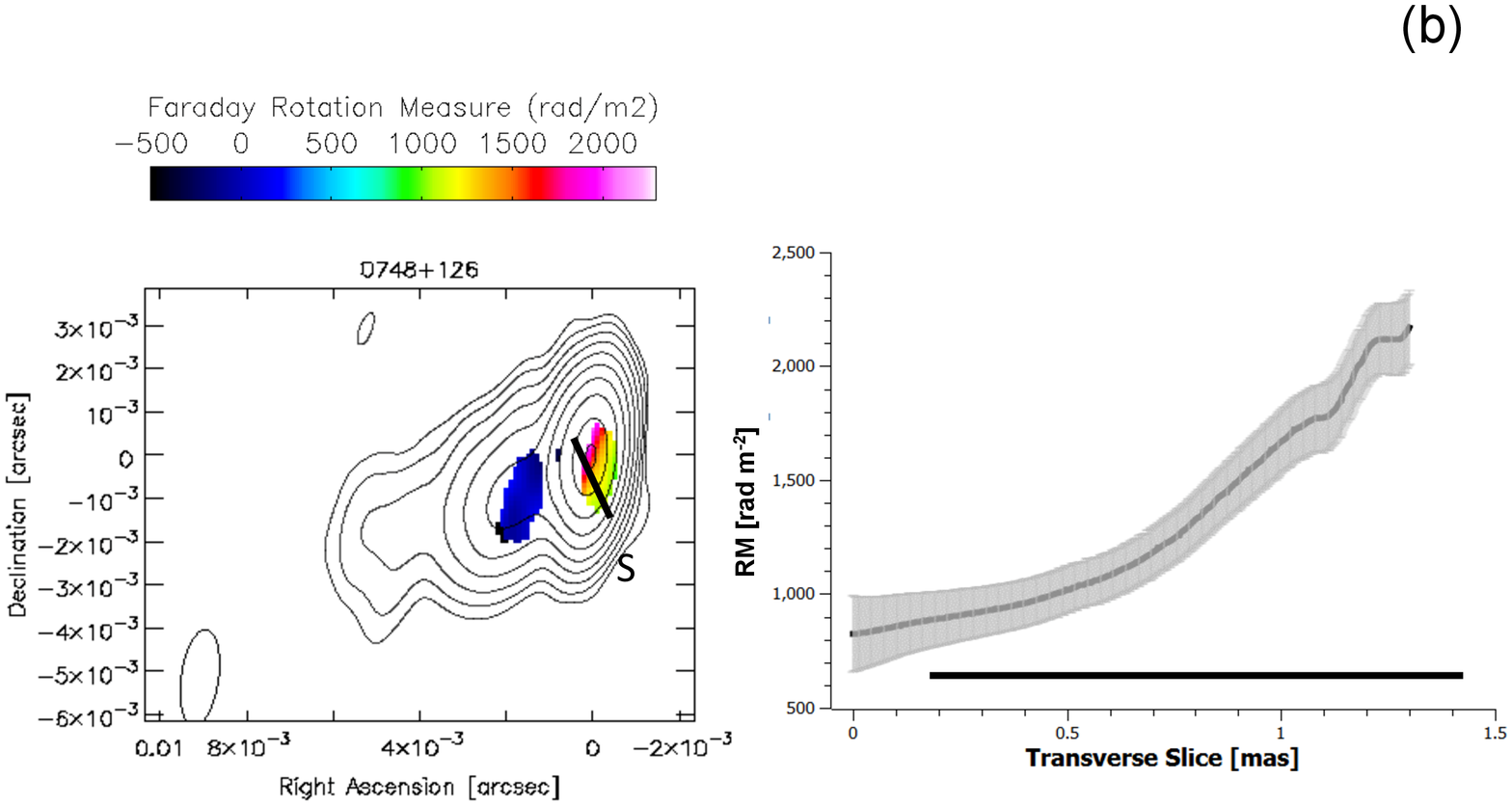}
\end{center}
\caption{8.1-GHz intensity maps of (a) 0355+508 and (b) 0748+126
with the corresponding
RM distributions superposed (left panels).  The lines drawn across the 
RM distributions show the locations of the RM slices shown in the 
corresponding right-hand panels; the letter ``S'' at one end of these 
lines marks the side corresponding to the starting point for the slice.
The bold horizontal bars shown with the slice profiles indicate the beam 
full widths at half maximum in the direction of the slices, which are 
(a) 1.0~mas and (b) 1.3~mas. }
\label{fig:rmmaps8}
\end{figure*}

\section{Observations }

We consider here both new analyses of previously published Faraday RM
maps and RM images published here for the first time.  In all cases, 
the observations were obtained on the NRAO Very Long Baseline Array.
We have applied the error estimation formula of Hovatta 
et al. (2012) to determine the uncertainties in the polarization angles 
in individual pixels. 

The observed Faraday rotation occurs predominantly in two
locations: in the immediate vicinity of the AGN and in our Galaxy. The
latter contribution must be estimated and removed if we wish to isolate
Faraday rotation occurring in the vicinity of the AGN itself.  The effect 
of the integrated (Galactic) RM is usually small, but can be substantial 
for some sources, e.g. those lying near the plane of the Galaxy.
We used various integrated RM measurements obtained using the Very Large
Array, indicated in Table~1, to remove the effect of the Galactic RM from 
the observed polarization angles, when significant, before making our RM maps.

When required, we corrected for any significant relative
shifts between the polarization angle images used to make the RM maps.
We determined these relative shifts using the cross-correlation approach
of Croke \& Gabuzda (2008); the shifts were tested by making 
spectral-index maps taking into account the relative shifts,
to ensure that they did not show any spurious features due to residual
misalignment between the maps.  

The maps were made with natural weighting.  In a number of cases when the 
beams obtained were fairly elongated, we also made versions of the RM
maps using circular beams with areas (roughly) equal to those of the intrinsic
elliptical beams, to test the robustness of gradients detected in the
original images.

\subsection{5--15~GHz, February 1997}

We consider the RM maps published by Gabuzda et al. (2004), as well as
several RM maps for other objects for which data were taken in the same
set of observations.  These observations focused on objects in the sample 
of 34 BL Lac objects defined by K\"uhr \& Schmidt (1990). 
The observations were carried out on 9th February 1997 at 5, 8.4 and
15~GHz. The procedures used to
calibrate the data and construct the published RM maps are described by
Gabuzda et al. (2004). When the original RM maps were published, no
attempt to check the alignment of the polarization angle images at
the three frequencies was made, and we correct this here. Gabuzda et al.
(2004) removed the Galactic RM values from their RM maps; we have done
this for the new RM maps analyzed here as well.

\subsection{5--15~GHz, April 1997}

Six compact BL Lac objects were observed with the VLBA on 6th April 1997, 
simultaneously at 22.2, 15.3, 8.4 and 5.0~GHz; we consider here only the 
lower three frequencies, since the difference in resolution between the
22-GHz and 5-GHz data is quite large. The procedures used to calibrate 
the data are described by Reynolds et al. (2001) and Gabuzda \& Chernetskii 
(2003).  The procedures used to construct the RM map for 1219+285 
considered here were the same as those used by Gabuzda et al. (2004). We 
have taken into
account the correct alignment between the polarization angle images at the
different frequencies in the RM map presented here. We have not attempted
to correct for the effect of Galactic Faraday rotation, as the integrated
RM is negligible in this case ($-1\pm 3$~rad/m$^2$; Rusk 1988). 

\subsection{8.1--15.2~GHz, June 2001}

Zavala \& Taylor (2004) present 15.2-GHz total intensity maps with 
superposed polarization sticks and Faraday RM maps based on 7 well-spaced
frequencies between 8.1 and 15.2~GHz for 17 AGN. 
These observations were carried out on 20th June 2001 at 8.1, 8.2, 8.4,
8.6, 12.1, 12.6 and 15.2~GHz.  The procedures used to
calibrate the data and construct the published RM maps are described by
Zavala \& Taylor (2004).  The final fully calibrated and self-calibrated
visibility data were kindly provided by R. Zavala, and it is these data
that we used to make the RM images analyzed here.  

A visual inspection of the published RM images indicated the possible 
presence of transverse RM gradients across the jets of 
0355+508 and 0748+126, as well as  1749+096, whose transverse RM
gradient was previously reported by Gabuzda et al. (2008). 
We made stokes $Q$ and $U$ maps in AIPS for each of these two
sources at each of the frequencies, 
ensuring that the image parameters (cell size, image
size, beam parameters) were the same for all 7 frequencies, and 
taking the errors to be given by the approach of Hovatta et al. (2012). 
We then used these to obtain matched-resolution polarization-angle (``PANG'') 
and polarization-angle noise (``PANGN'') images, which were used to
construct RM maps after removing the effect of integrated (Galactic)
Faraday rotation when significant.  These essentially reproduced the 
RM maps published by Zavala \& Taylor (2004), but with the more
conservative errors of Hovatta et al. (2012) and with the Galactic
RM values removed. 

Note that the procedure described above for matching the resolutions
of the images at the different frequencies was different from the approach 
adopted by Zavala \& Taylor (2004), who applied tapers to the 12 and
15~GHz data used to produce the polarization-angle maps so as to approximate
the 8-GHz resolution, and then used a restoring beam matched to the
8-GHz beam. Although the procedure used by Zavala \& Taylor (2004) is
formally more correct, we did not find any significant differences between
our RM maps and those of Zavala \& Taylor (2004); it is likely that any
differences in the polarization maps due to the application of these
different approaches appear at lower flux and polarization levels than 
those contributing to the RM images.

No mention of  
image alignment is made in the original publication, and the published
spectral-index maps suggest that no image alignment was done (they
often show a band of seemingly optically thin emission on the side of the
core opposite  the jet); we have ensured correct alignment of the 
polarization-angle images in our analysis.  

\subsection{4.6--15.1~GHz, August 2003, March 2004 and September 2004}

We consider here the published RM maps of Gabuzda et al. (2008), as well
as RM maps for several other objects for which data were taken in the
same sets of observations.
The observations were carried out at 4.6, 5.1, 7.9, 8.9, 12.9 and
15.4~GHz on 22nd August 2003, 22nd March 2004 and 10th September 2004.
The procedures used to calibrate the data and construct the published
RM maps are described by Gabuzda et al. (2008).
When the original RM maps were published, no
attempt to check the alignment of the polarization angle images at
the three frequencies was made, and we address this here.  Gabuzda et al.
(2008) removed the Galactic RM values from their RM maps; we have done
this for the new RM maps analyzed here as well.

\section{Results}

The source names, redshifts, optical identifications, pc/mas values and 
integrated rotation measures are summarized
in Table~1.  The pc/mas were determined assumed a cosmology with $H_o = 
71$~km\,s$^{-1}$Mpc$^{-1}$, $\Omega_{\Lambda} = 0.73$ and $\Omega_{m} = 
0.27$; the redshifts and pc/mas values were taken from the MOJAVE project 
website when available (http://www.physics.purdue.edu/MOJAVE/). 

RM maps together with slices in regions of detected transverse RM
gradients are shown in Figs.~\ref{fig:rmmaps5-1}--\ref{fig:rmmaps8}. 
The frequencies, peaks and bottom contours of the intensity maps shown in
these figures are given in Table~2; in all cases, the contour levels
increase in steps of a factor of two. The ranges of the RM maps are 
indicated by the colour wedges shown with the maps.
For consistency, in each case, the RM slices were taken in the clockwise 
direction relative to the base of the 
jet (located upstream from the observed core). The lines
drawn across the RM distributions show the locations of the slices; the 
letter ``S'' at one end of these lines marks the side corresponding to
the starting point for the slice (a slice distance of 0). The
slices were taken in regions where the significance of the RM gradients
was highest. The
full width at half maximum of the beam in the direction of the slice
is shown together with the slices by a bold bar.

The statistical significance of transverse gradients detected in our
RM maps are summarized in Table~3. When plotting the slices in 
Figs.~\ref{fig:rmmaps5-1}--\ref{fig:rmmaps8} and finding the difference 
between the RM values at two ends of a gradient, $\Delta$RM, we did not 
include uncertainty 
in the polarization angles due to EVPA calibration uncertainty; this 
is appropriate, since EVPA calibration uncertainty affects all 
polarization angles for the frequency in question equally, and so 
cannot introduce spurious RM gradients, as is discussed by Mahmud 
et al. (2009) and Hovatta et al.  (2012).  The contribution of uncertainty
in the polarization angles due to residual D-terms described by Hovatta
et al. (2012) has been included. Hovatta et al. (2012) defined the
uncertainty in $\Delta$RM to be the largest RM uncertainty at the edge 
of the jet; instead, we have taken the uncertainty $\Delta$RM to be 
the sum of the two RM uncertainties added in quadrature. While not
strictly correct mathematically due to the effects of convolution,
this at least takes into account the uncertainties at both ends of the
slice, and is more conservative than the approach of Hovatta et al.
(2012). In most cases, the RM values used to
calculate $\Delta$RM are located at or near the ends of the slices
shown; in some cases, we used values somewhat farther from the ends
of the slices because the increase in the uncertainties at the slice
ends appreciably reduced the statistical significance of the RM 
difference. In all cases, the RM slices showed monotonic changes in
the RM across the slices right to the slice ends.

Note that we do not reproduce
the RM maps previously published by Gabuzda et al.  (2004, 2008) here,
as an initial analysis of the transverse RM structure was already carried 
out in those papers; however, 
our results for those sources are included in Table~3.

Results for each of the AGN considered here are summarized briefly below.
We took a transverse RM gradient to be 
across the core if it was located within the 50\% intensity contour; 
otherwise, we took the gradient to be across the jet. 

\subsection{5--15~GHz, February 1997}

These RM maps are based on simultaneous VLBA observations at 15, 8 and
5~GHz.

\smallskip
\noindent
{\bf 0745+241.} 
This RM map was originally published by Gabuzda et al.  (2004). Our results 
confirm that the previously reported RM gradient is monotonic; the 
significance of this transverse gradient is about $4\sigma$.

\smallskip
\noindent
{\bf 0820+225.}
This RM map was originally published by Gabuzda et al.  (2004). Our results 
confirm that the previously reported RM gradient is monotonic; the significance 
of this transverse gradient is $3.4\sigma$.

\smallskip
\noindent
{\bf 0823+033.}
This RM map, shown in Fig.~\ref{fig:rmmaps5-1}b, has not been published 
previously. It shows a monotonic transverse RM gradient with a significance 
of about $4\sigma$.

\smallskip
\noindent
{\bf 1652+398.}
This RM map was originally published by Gabuzda et al.  (2004). Our results 
confirm 
that the previously reported RM gradient is monotonic; the significance of 
the transverse gradient is $4.8\sigma$. Note that a transverse RM gradient with
the same direction on somewhat larger scales has been reported by Croke et al.
(2010).

\smallskip
\noindent
{\bf 1807+698.}
This RM map was originally published by Gabuzda et al.  (2004). Our results 
confirm that the previously reported RM gradient is monotonic; the 
significance of the transverse gradient is $3.5\sigma$.

\smallskip
\noindent
{\bf 2007+777.}
This RM map, shown in Fig.~\ref{fig:rmmaps5-1}d, has not been published 
previously. It shows a monotonic transverse RM gradient with a significance 
of $5.5\sigma$.

\subsection{5--15~GHz, April 1997}

\noindent
{\bf 1219+285.} 
This RM map, shown in Fig.~\ref{fig:rmmaps5-1}c, has not been published 
previously. It shows a monotonic transverse RM gradient with a significance 
of about $3\sigma$.

\subsection{8.1--15.2~GHz, June 2001}

RM maps of these sources based on the same visibility data but with somewhat
different weighting were originally published by Zavala \& Taylor (2004). They 
were chosen for our analysis because possible RM gradients across the jet 
structure are visible by eye in the RM maps 
of Zavala \& Taylor (2004). 

\smallskip
\noindent
{\bf 0355+508.} 
Our map in Fig.~\ref{fig:rmmaps8}a shows a monotonic transverse RM gradient 
across the core region, with a significance of $4.2\sigma$.  The 
orientation of the beam in our image is roughly along the jet, making it 
possible that the significance of the gradient has been artificially increased 
by this effective averaging along the jet. The RM gradient remains visible 
when the RM map is made using a circular beam, with a significance of 
about $3\sigma$.

\smallskip
\noindent
{\bf 0748+126.} 
Our map in Fig.~\ref{fig:rmmaps8}b shows a transverse RM gradient across the
core region, with a significance of nearly $6\sigma$. 

\subsection{4.6--15.1~GHz, August 2003, March 2004 and September 2004}

\noindent
{\bf 0256+075.} 
An RM map made from the same visibility data was published by Mahmud \&
Gabuzda (2008), but without any error analysis.  Our map in 
Fig.~\ref{fig:rmmaps5-1}a shows a monotonic transverse RM gradient across the
inner jet, whose significance is $2.9\sigma$. We also made this RM map
using a circular beam; the RM gradient remains visible, and its 
significance increases to $3.2\sigma$.

\smallskip
\noindent
{\bf 0735+178.} 
An RM map made from the same visibility data was published by Gabuzda et al.
(2008).  Our analysis shows that the previously reported monotonic, transverse 
RM gradient across the jet has a significance of $5.5\sigma$. 

\smallskip
\noindent
{\bf 1156+295.} 
An RM map made from the same visibility data was published by Gabuzda et al.
(2008).  Our analysis shows that the previously reported monotonic, transverse 
RM gradient across the core region has a significance of $5.8\sigma$. 

\smallskip
\noindent
{\bf 1334$-$127.} 
Our map in Fig.~\ref{fig:rmmaps5-2} is the first RM map published for this
source in this frequency range. It shows a nearly monotonic transverse RM 
gradient across the core region, whose significance is $4.6\sigma$. 
We also constructed a version of this map with a circular beam, to ensure 
that the apparent transverse gradient was not an artefact of the low
declination of the source; the RM gradient remained visible in this
map, became monotonic, and had a similar significance, $5.0\sigma$.

\smallskip
\noindent
{\bf 1749+096.} 
An RM map made from the same visibility data was published by Gabuzda et al.
(2008). Our analysis shows that the previously reported monotonic, transverse 
RM gradient across the core has a significance is $5.5\sigma$. This gradient 
is also visible in the 8.1--15.2~GHz RM image of Zavala \& Taylor (2004).

\smallskip
\noindent
{\bf 2155$-$152.}
An RM map made from the same visibility data was published by Mahmud \&
Gabuzda (2008), but without any error analysis.  Mahmud \& Gabuzda (2008)
suggested the presence of two oppositely directed transverse RM gradients ---
one in the core region and one in the jet. Due to the rather different
RM ranges for the core region and jet, we show separate RM maps for these
two regions in Figs.~\ref{fig:rmmaps5-2}b,c. There appears to be a 
transverse RM gradient across the core, but it is difficult to estimate
its significance due to the elongated beam. A version of this map made
with a circular beam shows a clear, monotonic RM gradient with a 
significance of $5.0\sigma$.  The possible oppositely directed gradient 
further out in the jet reported by Mahmud et al. (2008) is not monotonic
in either the map made with the intrinsic beam or the circular beam,
and we therefore do not consider it to be convincing.

\begin{center}
\begin{table*}
\begin{tabular}{c|c|c|c|c|c}
\hline
\multicolumn{6}{c}{Table 3: Summary of transverse RM gradients}\\
Source & Location & RM$_1$ & RM$_2$ & $|\Delta$RM$|$ & Significance  \\ 
       &          & rad/m$^2$ & rad/m$^2$ & rad/m$^2$ & \\
0256+075 -- EB & Jet  & $-107\pm 69$ & $165\pm 63$  & $272\pm 93$  & $2.9\sigma$\\
0256+075 -- CB & Jet  & $-75\pm 46$ & $142\pm 49$  & $217\pm 67$  & $3.2\sigma$\\
0355+508 & Core & $-662\pm 109$& $-1283\pm 99$& $621\pm 147$& $4.2\sigma$\\
0735+178  & Jet  & $-408\pm 86$& $187\pm 63$   & $595\pm 107$& $5.5\sigma$\\
0745+241  & Jet  &$-99\pm 15$  & $34\pm 28$    & $133\pm 32$  & $4.2\sigma$\\
0748+126  & Core  & $824\pm 166$ & $2172\pm 160$ &$1348\pm230$ & $5.9\sigma$\\
0820+225  & Jet  &$26\pm 16$   & $140\pm 30$    &$114\pm 34$  & $3.4\sigma$\\
0823+033  & Jet  & $-93\pm 18$ & $-216\pm 25$  & $123\pm 31$ & $4.0\sigma$ \\
1156+295  & Core  & $117\pm31$ & $345\pm 24$   & $228\pm 39$ & $5.8\sigma$\\
1219+285  & Jet  & $-5\pm 19$  & $89\pm 25$    & $94\pm 31$  & $3.0\sigma$  \\
1334$-$127 -- EB  & Jet  & $-86\pm48$  & $-367\pm37$   & $281\pm 61$ & $4.6\sigma$\\
1334$-$127 -- CB & Jet & $-62\pm38$ & $-333\pm38$ & $271\pm 54$  & $5.0\sigma$\\
1652+398  & Jet  & $167\pm 35$ & $-41\pm 26$   & $167\pm 35$  & $4.8\sigma$\\
1749+096 & Core & $53\pm 48$  & $394\pm 40$  & $341\pm 62$  & $5.5\sigma$\\
1807+698  & Jet  & $225\pm 49$ & $501\pm 62$   & $276\pm 79$  & $3.5\sigma$\\
2007+777  & Core & $-60\pm 28$ & $-241\pm 18$  & $181\pm 33$ & $5.5\sigma$\\
2155$-$152 -- CB  & Core  & $-566\pm13$ & $-303\pm50$ & $263\pm 52$  & $5.0\sigma$\\
\\ \hline
\multicolumn{6}{l}{EB denotes elliptical beam and CB a circular beam of
roughly the same area as the intrinsic elliptical beam.}\\
\end{tabular}
\end{table*}
\end{center}

\section{Discussion}  

\subsection{Reliability and Significance of the Transverse RM Gradients}

Table~3 gives a summary of the transverse Faraday rotation measure gradients 
detected in the images presented here. The statistical significances of
these gradients range from $3\sigma$ to nearly $6\sigma$. This
indicates that these transverse RM gradients are not likely to be spurious 
(i.e., due to inadequacy of the $uv$ coverage and noise): the Monte Carlo 
simulations of Hovatta et al. (2012) and
Murphy \& Gabuzda (2013) have shown that spurious gradients at the
$3\sigma$ level should arise for 4--6-frequency VLBA observations in the 
frequency interval considered here with a probability of no more than about
1\%, with this probability being even lower for monotonic gradients 
encompassing differences of greater than $3\sigma$. Although the occurrence
of spurious gradients rises substantially for VLBA observations of 
low-declination sources, the significances of the gradients observed
across the cores/inner jets of 1334$-$127 and 2155$-$152 ($4.5-5\sigma$) 
are high enough to make it quite improbable that these gradients are spurious.


Sign changes are observed in the transverse RM gradients detected in
0256+075, 0735+178, 0745+241 and 1652+398. This strengthens the case that 
these gradients are due to helical (or toroidal) magnetic fields associated 
with these jets, since a sign change cannot be caused by gradients in the
electron density, and must be associated with a change in the direction of 
the line-of-sight magnetic field. 

Note, however, that the absence of a sign change in the transverse RM profile
does not rule out the possibility that a transverse gradient is due
to a helical or toroidal field component in the region of Faraday
rotation, since gradients encompassing only one sign can be observed
for some combinations of helical pitch angle and viewing angle.

\subsection{Core-Region Transverse RM Gradients}

In the standard theoretical picture, the VLBI ``core'' represents  
the ``photosphere'' at the base of the jet, where the optical depth is
roughly unity. However, in images with the resolution of those presented
here, the observed core is actually a blend of this (partially) optically 
thick region and optically thin regions in the innermost jet. Since these 
optically thin regions are characterized by  degrees of linear polarization 
that are typically a factor of 10 or more higher than at the optically thick
base of the jet, they likely dominate the
overall observed ``core'' polarization in many cases. Therefore, we have 
supposed that the polarization angles observed in the core are most likely
orthogonal to the local magnetic field, as expected for predominantly
optically thin regions.  We have not observed 
any sudden jumps in polarization
position angle by roughly $90^{\circ}$ suggesting the presence of optically 
thick--thin transitions in the cores in our frequency ranges at the
observing epochs.

This picture of the observed VLBI core at centimeter wavelengths corresponding
to a mixture of optically thick and thin regions, with the observed core 
polarization contributed predominantly by optically thin regions, also 
impacts our interpretation of the core-region Faraday rotation measures.
Monotonic transverse RM gradients with significances of at least $3\sigma$
are observed across the core regions of 
0355+508, 0748+126, 1156+295, 1749+096, 2007+777 and 2155$-$152. 

The simplest approach to interpreting 
these gradients is to treat them in the same way as transverse gradients
observed outside the core region, in the jet. While the simulations of 
Broderick \& McKinney (2010) show that relativistic and optical depth effects
can sometimes give rise to non-monotonic transverse RM gradients in core
regions containing helical magnetic fields, there are also cases when these
helical fields give rise to monotonic RM gradients, as they would in a fully
optically thin region. In addition, most of the non-monotonic behaviour
that can arise will be smoothed by convolution with a typical 
centimeter-wavelength VLBA beam [see, for example, the lower right panel in
Fig.~8 of Broderick \& McKinney (2010)]. 
Therefore, when a smooth, monotonic, 
statistically significant transverse RM gradient is observed across the 
core region, it is reasonable to interpret this as evidence for 
helical/toroidal fields in this region (i.e., in the innermost jet).
 
\section{Conclusion}

We have confirmed previous reports of transverse Faraday RM gradients across
the jet structures of 7 AGNs (4.6--15.4~GHz data; Gabuzda et al. 2004, 2008), 
applying the error-estimation approach of Hovatta et al. (2012) and ensuring
correct alignment of the polarization-angle images at the different frequencies
used to construct the RM maps. Our analysis indicates all of these gradients
to be monotonic and to have significances of at least $3.4\sigma$.

We have also investigated the reality of transverse RM gradients visible 
in the previously published RM maps for 2 additional AGNs (8.1--15.2~GHz data; 
Zavala \& Taylor 2004), which are likewise monotonic and have significances
of at least $3\sigma$. 

Finally, we have reported new monotonic transverse Faraday RM gradients with
significances of $3\sigma$ or more in another 6 AGNs for which maps were 
not published previously in the refereed literature (4.6--15.4~GHz data). 

In all, the analysis carried out in this study has added 15 sources to the list 
of AGNs whose jet structures display monotonic transverse RM gradients with 
significances of at least $3\sigma$, based on the most up-to-date methods for
error estimation and image analysis. One reasonable interpretation of these
gradients is that they reflect the presence of helical or toroidal magnetic
fields, which form due to the combination of the rotation of the central black
hole and accretion disk and the jet outflow, and then travel outward with the 
jet material.  Four of these gradients encompass RM
values of both signs, strengthening the case that they are associated
with a toroidal magnetic-field component, since they cannot be explained
by gradients in the electron density.

It would be of interest to investigate systematic changes in the 
transverse RM gradients along the jet. Unfortunately, this is hindered
by the superposition of a more random RM component, presumably due to
turbulence in the medium surrounding the jet, through which the polarized
jet radiation passes. In the few cases when a transverse RM gradient is 
clearly observed over a range of distances from the jet base spanning more
than a beamwidth, as in 0820+225 (Gabuzda et al. 2004), 3C273 (e.g. Hovatta 
et al. 2012) and 3C380 (Gabuzda et al. 2014a), the gradient appears to be
fairly uniform in the region where it is observed. More detailed analyses
of variations in the RM gradients along the jet will be carried out in a
separate study. We are also currently
engaged in a project to investigate transverse RM gradients farther from 
the jet base than those considered here, using longer-wavelength VLBA and 
VLA data. 

\section{Acknowledgements}

We thank Robert Zavala for providing the version of the AIPS
task RM used for this work.  We are also grateful to the anonymous referee 
for his or her quick, thorough and helpful reviews.
The National Radio Astronomy Observatory is a facility of the National 
Science Foundation operated under cooperative agreement by Associated 
Universities, Inc.

\end{document}